# Logic gates at the surface code threshold: Superconducting qubits poised for fault-tolerant quantum computing

R. Barends,[1,*] J. Kelly,[1,*] A. Megrant,[1] A. Veitia,[2] D. Sank,[1] E. Jeffrey,[1] T. C. White,[1] J. Mutus,[1] A. G. Fowler,[1,3] B. Campbell,[1] Y. Chen,[1] Z. Chen,[1] B. Chiaro,[1] A. Dunsworth,[1] C. Neill,[1] P. O'Malley,[1] P. Roushan,[1] A. Vainsencher,[1] J. Wenner,[1] A. N. Korotkov,[2] A. N. Cleland,[1] and John M. Martinis[1]

[1]*Department of Physics, University of California, Santa Barbara, CA 93106, USA*
[2]*Department of Electrical Engineering, University of California, Riverside, CA 92521, USA*
[3]*Centre for Quantum Computation and Communication Technology, School of Physics, The University of Melbourne, Victoria 3010, Australia*

**A quantum computer can solve hard problems - such as prime factoring[1,2], database searching[3,4], and quantum simulation[5] - at the cost of needing to protect fragile quantum states from error. Quantum error correction[6] provides this protection, by distributing a logical state among many physical qubits via quantum entanglement. Superconductivity is an appealing platform, as it allows for constructing large quantum circuits, and is compatible with microfabrication. For superconducting qubits the surface code[7] is a natural choice for error correction, as it uses only nearest-neighbour coupling and rapidly-cycled entangling gates. The gate fidelity requirements are modest: The per-step fidelity threshold is only about 99%. Here, we demonstrate a universal set of logic gates in a superconducting multi-qubit processor, achieving an average single-qubit gate fidelity of 99.92% and a two-qubit gate fidelity up to 99.4%. This places Josephson quantum computing at the fault-tolerant threshold for surface code error correction. Our quantum processor is a first step towards the surface code, using five qubits arranged in a linear array with nearest-neighbour coupling. As a further demonstration, we construct a five-qubit Greenberger-Horne-Zeilinger (GHZ) state[8,9] using the complete circuit and full set of gates. The results demonstrate that Josephson quantum computing is a high-fidelity technology, with a clear path to scaling up to large-scale, fault-tolerant quantum circuits.**

The high fidelity performance we demonstrate here is achieved through a combination of highly coherent qubits, a straightforward interconnection architecture, and a novel implementation of the two-qubit controlled-phase (CZ) entangling gate. The CZ gate uses a fast but adiabatic frequency tuning of the qubits[10], which is easily adjusted yet minimises decoherence and leakage from the computational basis [Martinis, J., *et al.*, in preparation]. We note that previous demonstrations of two-qubit gates achieving better than 99% fidelity used fixed-frequency qubits: Systems based on nuclear magnetic resonance and ion traps have shown two-qubit gates with fidelities of 99.5%[11] and 99.3%[12]. Here, the tuneable nature of the qubits and their entangling gates provides, remarkably, both high fidelity and fast control.

Superconducting integrated circuits give flexibility in building quantum systems due to the macroscopic nature of the electron condensate. As shown in Fig. 1, we have designed a processor consisting of five Xmon qubits with nearest-neighbour coupling, arranged in a linear array. The cross-shaped qubit[14] offers a nodal approach to connectivity while maintaining a high level of coherence (see Supplementary Information for decoherence times). Here, the four legs of the cross allow for a natural segmentation of the design into coupling, control and readout. We chose a modest inter-qubit capacitive coupling strength of $g/2\pi = 30$ MHz and use alternating qubit idle frequencies of 5.5 and 4.7 GHz, enabling a CZ gate in 40 ns when two qubits are brought near resonance, while minimising the effective coupling to 0.3 MHz when the qubits are at their idle points. Rotations around the X and Y axes in the Bloch sphere representation are performed using pulses on the microwave (XY) line, while Z axis rotations are achieved by a flux-bias current on the frequency-control (Z) line. We use dispersive measurement[15] where each qubit is coupled to a readout resonator, each with a different resonance frequency, enabling simultaneous readout using frequency-domain multiplexing through a single coplanar waveguide[16][Mutus, J., *et al.*, in preparation]. The modularity of this architecture makes it straightforward to integrate more qubits in the circuit.

We characterise our gate fidelities using Clifford-based randomised benchmarking[11,17,18]. The Clifford group is a set of rotations that evenly samples the Hilbert space, thus averaging across errors. For the single-qubit case the Cliffords are comprised of $\pi$, $\pi/2$ and $2\pi/3$ rotations, see Supplementary Information. In randomised benchmarking, a logic gate is characterised by measuring its performance when interleaved with many random sequences of gates, making the measured fidelity resilient to state preparation and measurement (SPAM) errors. We perform a control experiment on a ground-state qubit by: I) generating a random sequence of $m$ Cliffords, II) appending the unique recovery Clifford ($C_\mathrm{r}$) that makes the ideal sequence the identity, and III) averaging the experimental sequence fidelity, the final ground state population, over $k$ different sequences[18,19]. The resulting reference sequence fidelity $F_\mathrm{ref}$ is fit to $F_\mathrm{ref} = Ap_\mathrm{ref}{}^m + B$, where $p_\mathrm{ref}$ is the sequence decay, and state preparation and measurement errors are captured in the parameters $A$ and $B$. The average error per Clifford of the reference is given by $r_\mathrm{ref} = (1-p_\mathrm{ref})(d-1)/d$, with $d = 2^{N_\mathrm{qubits}}$. We then measure the fidelity of a specific gate by interleaving this test gate with $m$ random Cliffords and performing the same measurement. The sequence decay $p_\mathrm{gate}$ then gives the gate error $r_\mathrm{gate} = (1 - p_\mathrm{gate}/p_\mathrm{ref})(d-1)/d$.

The benchmarking results for the single-qubit gates are shown in Fig. 2. We generate the Cliffords using microwave pulses, from a basis set of $\pi$ and $\pi/2$ rotations around the X and Y axes (Supplementary Information). We benchmark X and Y axis $\pi$ and $\pi/2$ rotations, the Hadamard gate (implemented with Y/2 followed by X), and Z axis rotations using pulses on the frequency control line. From the data in Fig. 2 we extract the individual gate fidelities listed in the legend. We find an average fidelity of 99.92 % over all gates and qubits (Supplementary Information). The best fidelities are achieved by optimising the pulse amplitude and frequency, and minimising 2-state leakage[20] [Kelly, J., *et al.*, in preparation].

We have also measured the performance when simultaneously operating nearest or next-nearest qubits[21], operating them at dissimilar idle frequencies to minimise coupling. The fidelities are essentially unchanged, with small added errors $< 2 \cdot 10^{-4}$ (Supplementary Information), showing a high degree of addressability for this architecture.

The two-qubit CZ gate is implemented by tuning one qubit in frequency along a "fast adiabatic" trajectory which takes the two-qubit $|11\rangle$ state close to the avoided-level crossing with the $|02\rangle$ state, yielding a state-dependent relative phase shift. This implementation is the natural choice for weakly anharmonic, frequency-tunable qubits, as the other computational states are left unchanged[8,22,23]. Having the CZ gate adiabatic as well as fast is advantageous. An adiabatic trajectory is easily optimised and allows for exponentially suppressing leakage into the non-computational $|02\rangle$-state with gate duration. Having a fast CZ gate minimises the accumulation of errors from decoherence and unwanted entanglement with other circuit elements, favourable for fault-tolerance.

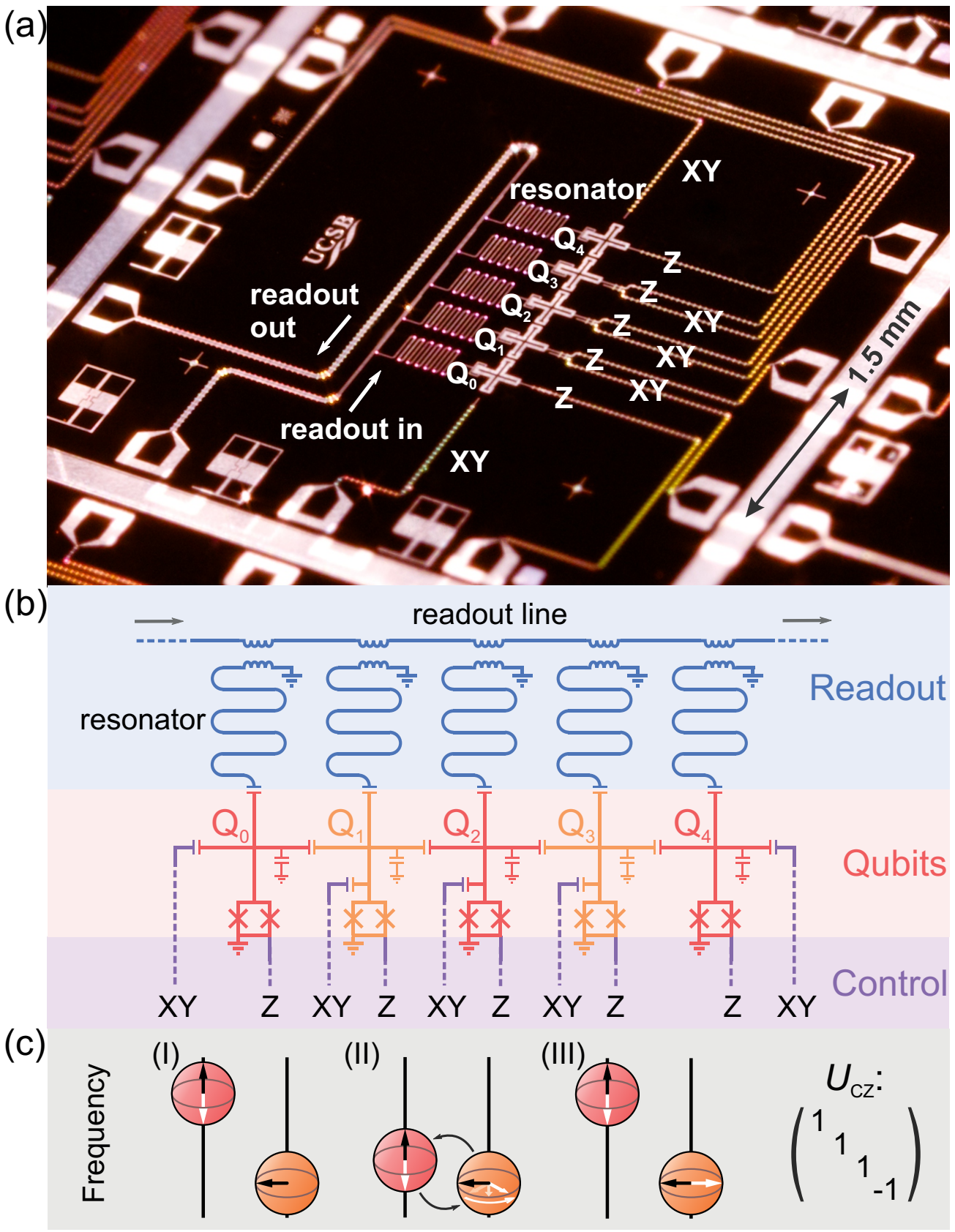

FIG. 1: **Architecture.** (a) Optical image of the integrated Josephson quantum processor, consisting of Al (dark) on sapphire (light). The five cross-shaped devices are the Xmon variant of the transmon qubit[13], labelled $Q_0 - Q_4$, placed in a linear array. To the left of the qubits are five meandering coplanar waveguide resonators used for individual state readout. Control wiring is brought in from the contact pads at the edge of the chip, ending at the right of the qubits. (b) Circuit diagram. Our architecture employs direct, nearest-neighbour coupling of the qubits (red/orange), made possible by the nodal connectivity of the Xmon qubit. Using a single readout line, each qubit can be measured using frequency-domain multiplexing (blue). Individual qubits are driven through capacitively-coupled microwave control lines (XY), and frequency control is achieved through inductively-coupled dc lines (Z) (purple). (c) Schematic representation of an entangling operation using a controlled-Z gate with unitary representation $U_{\mathrm{CZ}}$: (I) Qubits at rest, at distinct frequencies with minimal interaction. (II) When brought near resonance, the state-dependent frequency shift brings about a rotation conditional on the qubit states. (III) Qubits are returned to their rest frequency.

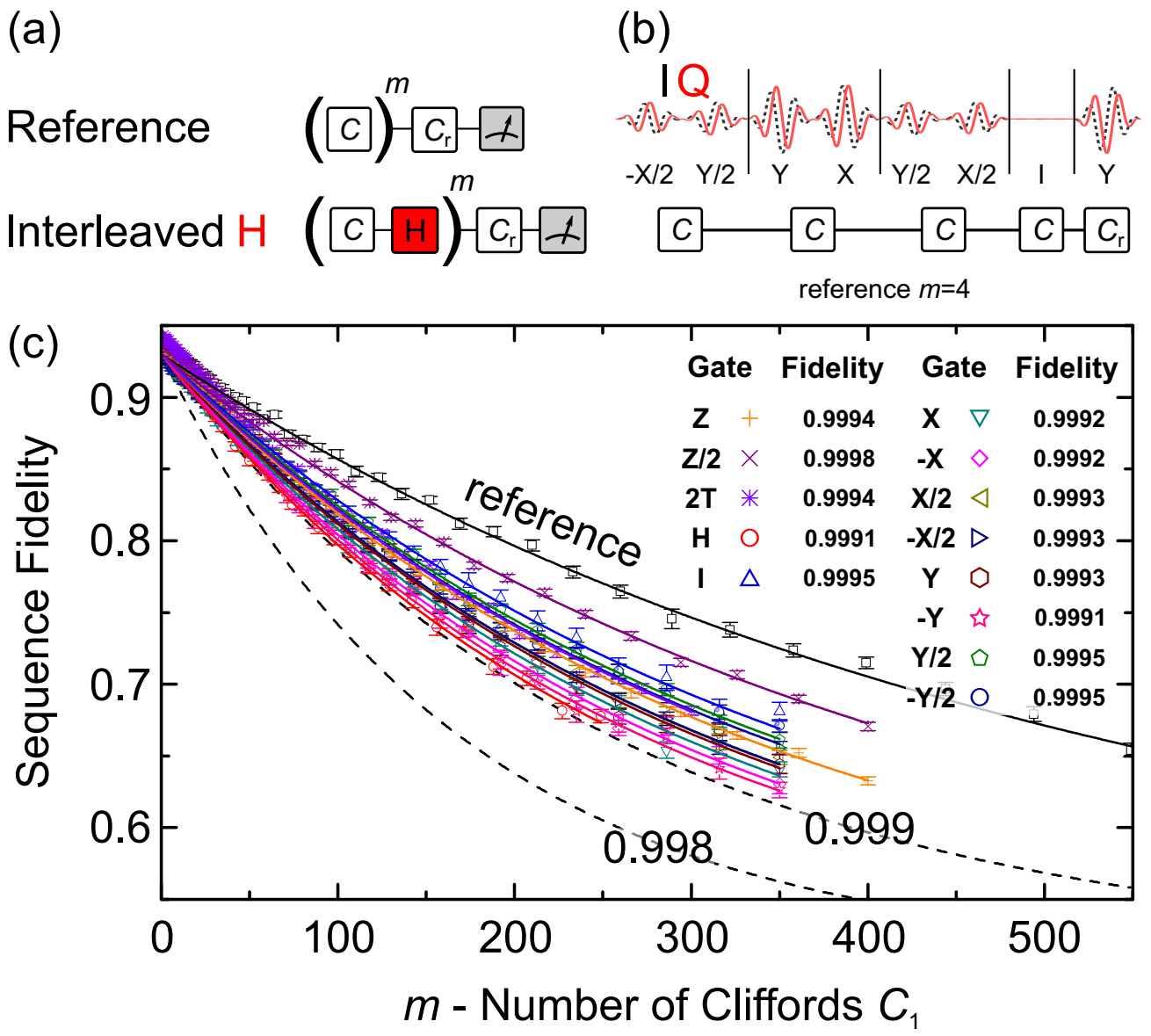

FIG. 2: **Single qubit randomised benchmarking.** (a) A reference experiment is performed by generating a sequence of $m$ random Cliffords, which are inverted by the recovery Clifford $C_r$. A specific gate ($H$) is tested using a sequence that interleaves $H$ with $m$ random Cliffords. The difference between interleaved and reference decay gives the gate fidelity. (b) Representative pulse sequence for a set of four Cliffords and their recovery, generated with $\pi$ and $\pi/2$ rotations about $X$ and $Y$, displaying both the real (I) and imaginary (Q) microwave pulse envelopes before up-conversion by quadrature mixing to the qubit frequency. (c) Randomised benchmarking measurement for the set of single-qubit gates for qubit $Q_2$, plotting reference and gate fidelities as a function of the sequence length $m$; the fidelity for each value of $m$ was measured for $k = 40$ different sequences. The fit to the reference set yields an average error per Clifford of $r_{\mathrm{ref}} = 0.0011$, consistent with an average gate fidelity of $1 - r_{\mathrm{ref}}/1.875 = 0.9994$ (Supplementary Information). The dashed lines indicate the thresholds for exceeding gate fidelities of 0.998 and 0.999. The fidelities for each of the single-qubit gates are tabulated in the legend, we find that all gates have fidelities greater than 0.999. Standard deviations are typically $5 \cdot 10^{-5}$.

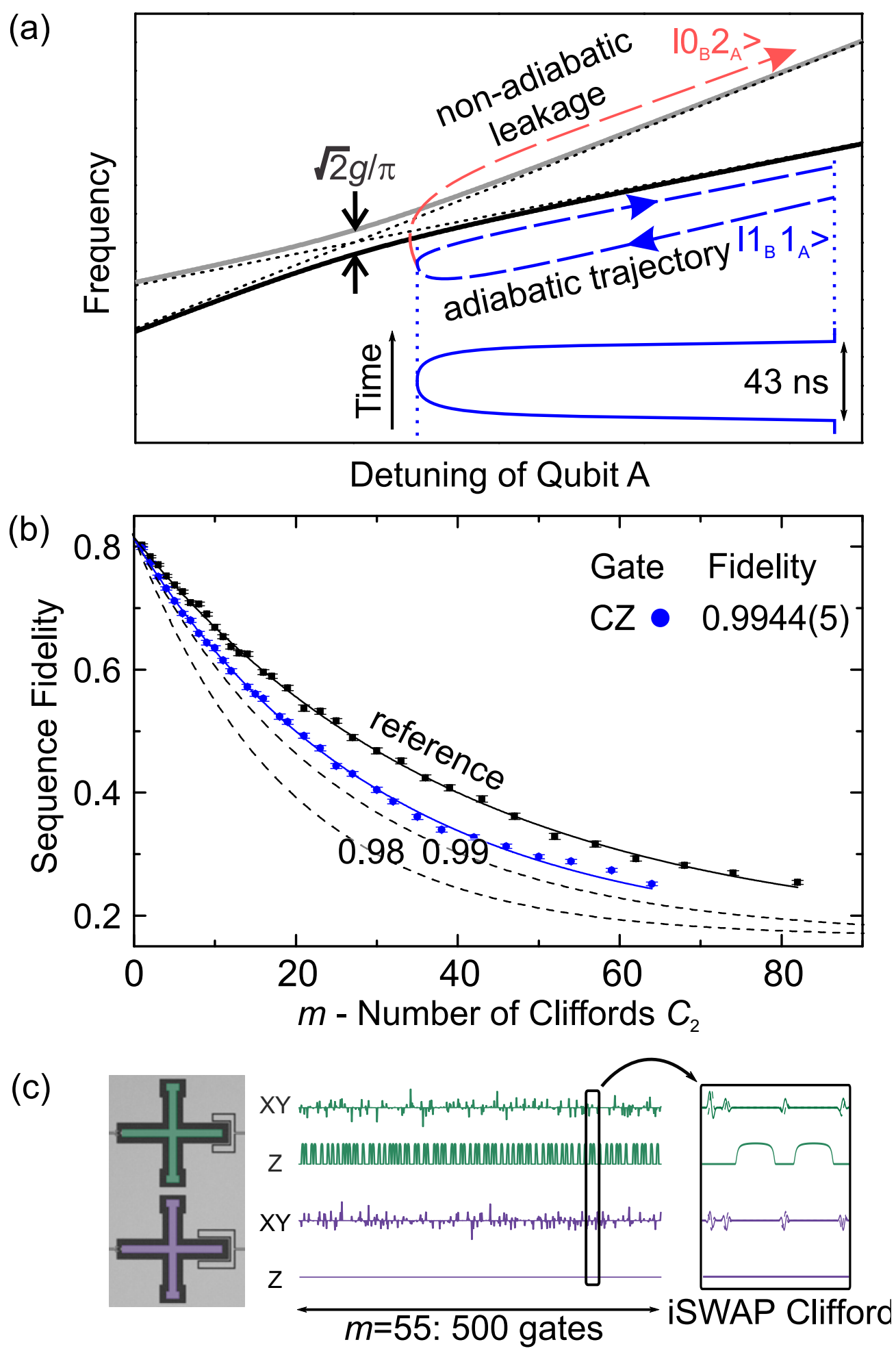

FIG. 3: **CZ gate physics and randomised benchmarking results.** (a) We use the the $|1_{\mathrm{B}}1_{\mathrm{A}}\rangle$ and $|0_{\mathrm{B}}2_{\mathrm{A}}\rangle$ avoided level crossing to implement a high-fidelity CZ gate, with the fast adiabatic tuning of qubit A giving a selective $\pi$ phase change of the $|1_{\mathrm{B}}1_{\mathrm{A}}\rangle$ state. The energy level diagram shows qubit A approaching and leaving the avoided level crossing from above (top, blue dashed line), following a fast (43 ns) yet effectively adiabatic trajectory (bottom, solid blue line). Unwanted state leakage from $|1_B1_A\rangle$ to $|0_B2_A\rangle$ (red dashed line) is minimised by adjusting the trajectory. (b) Randomised benchmarking data ($k = 100$) of the CZ gate for the qubit pair $Q_2$ and $Q_3$, using the two-qubit Clifford group $C_2$ (Supplementary Information); reference data in black ($r_{\mathrm{ref}} = 0.0189$), interleaved in blue ($r_{C_2+\mathrm{CZ}} = 0.0244$). Dashed lines indicate the thresholds for gate fidelities of 0.98 and 0.99. We find a CZ gate fidelity of $0.9944 \pm 0.0005$ (uncertainty from bootstrapping). (c) Coherent microwave (XY) and frequency (Z) control of the quantum state while performing a complex two-qubit algorithm; the sequence contains over 500 gates, corresponding to the characteristic reference decay of $m = 55$, and is over 7 $\mu$s long. The right panel shows an example Clifford from the iSWAP class, comprised of single qubit rotations and two CZ gates (Supplementary Information).

The benchmarking results of the CZ gate are shown in Fig. 3b. Similar to the single-qubit case, we generate sequences of two-qubit Cliffords to produce a reference curve, then interleave the CZ gate to extract the fidelity. An example pulse sequence for an $m = 55$ Clifford sequence is shown in Fig. 3c. We find a CZ gate fidelity of up to $99.44 \pm 0.05$ %, consistent with the average error per Clifford (Supplementary Information). We find fidelities between 99.0-99.44% on all four pairs of nearest-neighbour qubits (Supplementary Information). This comprises a clear demonstration of high-fidelity single- and two-qubit gates in a multi-qubit Josephson quantum processor. The two-qubit gate fidelity compares well with the highest values reported for other mature quantum systems: For nuclear magnetic resonance and ion traps, entangling gate fidelites are as high as 99.5% and 99.3%[11,12]. Importantly, we have verified by simulation that the experimentally obtained gate fidelities are at the threshold for surface code quantum error correction, see Supplementary Information.

We are optimistic that we can improve upon these gate fidelities with modest effort. The CZ gate fidelity is limited by three error mechanisms: Decoherence (55% of the total error), control error (24%), and state leakage (21%), see Supplementary Information. Decoherence can be suppressed with enhanced materials and optimised fabrication[24,25]. Imperfections in control arise primarily from reflections and stray inductances in wiring, and can be improved using conventional microwave techniques. Given the adiabatic nature of the CZ gate, 2-state leakage can be suppressed by slightly increasing the gate time [Martinis, J., *et al.*, in preparation].

We showcase the modularity of this set of quantum logic gates by constructing a maximally-entangled GHZ state across all five qubits in our processor, as shown in Fig. 4a. The $N$-qubit GHZ state $|\mathrm{GHZ}\rangle = (|0\rangle^{\otimes N} + |1\rangle^{\otimes N})/\sqrt{2}$ is constructed with single and two-qubit gates, using simultaneous control and readout of all qubits. This algorithm is shown in Fig. 4b, where the state is assembled by entangling one additional qubit at a time. The algorithm is highly sensitive to control error and decoherence on any computational element. We fully characterise the Bell and GHZ states by using quantum state tomography[9], where quadratic maximum likelihood estimation is used to extract each density matrix ($\rho$) from the measurement data, while satisfying the physical constraints that $\rho$ be Hermitian, unit trace, and positive semi-definite (Supplementary Information). The density matrices are plotted in the traditional cityscape style, and show significant elements only at the ideal locations. We find state fidelities $\mathrm{Tr}\left(\sqrt{\rho_{\mathrm{ideal}}}\rho\sqrt{\rho_{\mathrm{ideal}}}\right)$ of $99.5 \pm 0.4$ %, $96.0 \pm 0.5$ %, $86.3 \pm 0.5$ % and $81.7 \pm 0.5$ % for the $N = 2$ Bell state and $N = 3, 4, 5$ GHZ states. A GHZ state fidelity over 50 % satisfies the criterion for genuine entanglement[26]. It is interesting to note that the ratio of the off-diagonal to diagonal amplitudes $|\rho_{|0\rangle^{\otimes N},|1\rangle^{\otimes N}}|^2/\rho_{|0\rangle^{\otimes N},|0\rangle^{\otimes N}}\rho_{|1\rangle^{\otimes N},|1\rangle^{\otimes N}}$ have the values 0.99, 0.98, 0.99 and 0.99, suggesting that dephasing is small and/or uncorrelated. The five-qubit GHZ state is the largest multi-qubit entanglement demonstrated to date in the solid state[8,9], with state fidelity similar to results obtained in ion traps[27]. This demonstrates that complex quantum states can be constructed with high fidelity in a modular fashion, highlighting the potential for more intricate algorithms on this multi-purpose quantum processor.

We have shown single and two-qubit gates with fidelities at the fault-tolerant threshold for the surface code in an integrated circuit quantum processor. With this demonstration, Josephson quantum devices are now poised to explore fault-tolerant, multi-qubit computing. Extending the linear array of qubits to larger qubit numbers is straightforward, and generating a two-dimensional grid of qubits appears to be mostly a (significant) engineering challenge. In a separate experiment, we have demonstrated fast, high-fidelity qubit state measurement [Jeffrey, E., *et al.*, in preparation], in a design that can be seamlessly integrated with this architecture. The combination of high-fidelity logic, a multi-qubit architecture, and fast and accurate qubit readout provides the essential ingredients for a Josephson surface code quantum computer.

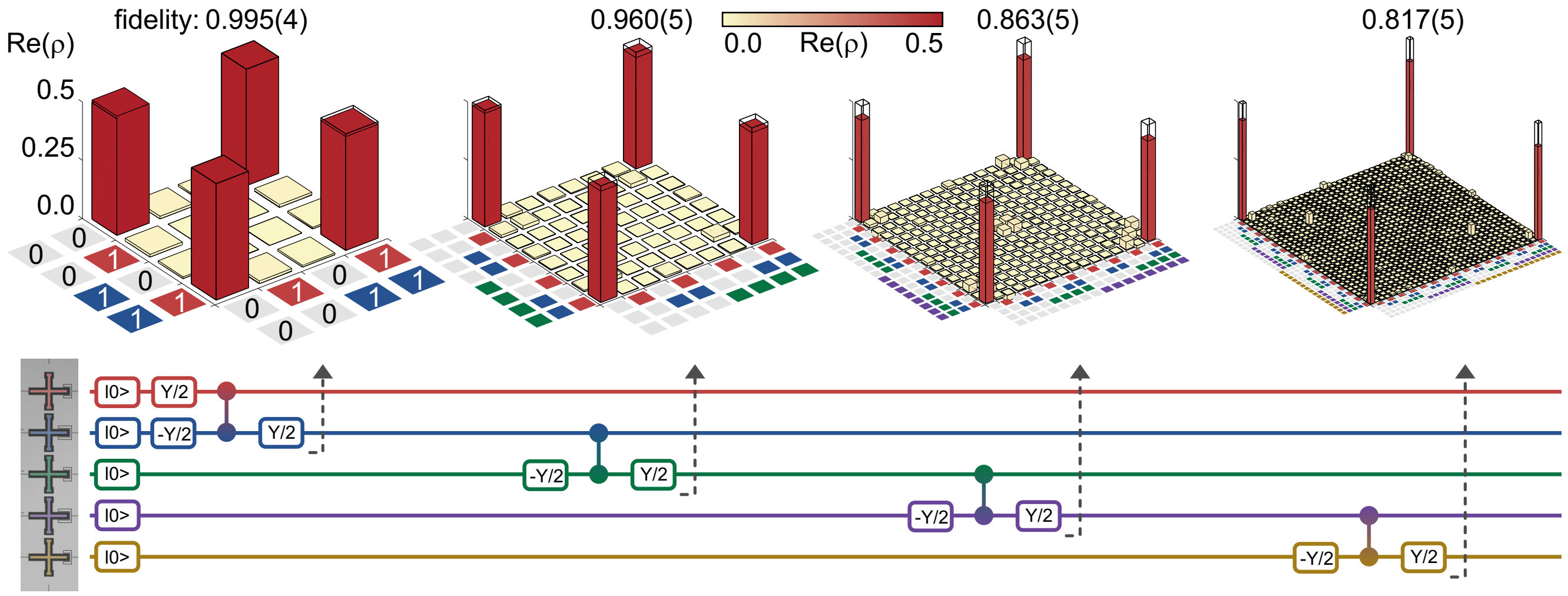

FIG. 4: **Quantum state tomography and generation of the GHZ state.** Top row: Real part of the density matrix $\rho$ for the $N = 2$ Bell state and the $N = 3$, 4 and 5 GHZ states, measured by quantum state tomography. Ideal density matrix elements are transparent, with value 0.5 at the four corners. Bottom row: Algorithm used to construct the states. See Supplementary Information for $\mathrm{Im}(\rho)$, the Pauli operator representation, and the full gate sequence, which includes Hahn spin-echo pulses.

**Acknowledgements** We thank F. Wilhelm, D. Egger, and J. Baselmans for helpful discussions. We are indebted to Erik Lucero for photography of the device. This work was supported by the Office of the Director of National Intelligence (ODNI), Intelligence Advanced Research Projects Activity (IARPA), through the Army Research Office grants W911NF-09-1-0375 and W911NF-10-1-0334. All statements of fact, opinion or conclusions contained herein are those of the authors and should not be construed as representing the official views or policies of IARPA, the ODNI, or the U.S. Government. Devices were made at the UC Santa Barbara Nanofabrication Facility, a part of the NSF-funded National Nanotechnology Infrastructure Network, and at the NanoStructures Cleanroom Facility.

**Author contributions** R.B. and J.K designed the sample, performed the experiment and analysed the data. J.K., A.E.M., and R.B. fabricated the sample. R.B., J.K., J.M.M., and A.N.C. co-wrote the manuscript. A.V. and A.N.K. provided assistance with randomised benchmarking. A.G.F. verified the experimental gate fidelities to be at the surface code threshold. All authors contributed to the fabrication process, experimental set-up and manuscript preparation.

**Additional information** The authors declare no competing financial interests. Supplementary information accompanies this paper on [weblink to be inserted by editor]. Reprints and permissions information is available at [weblink to be inserted by editor]. Correspondence and requests for materials should be addressed to R.B., J.K or J.M.M.

---

[*] These authors contributed equally to this work

# Supplementary Information

## DEVICE FUNDAMENTALS

### Fabrication

The devices are made in a process similar to the fabrication steps outlined in Ref. [1], with an important improvement: we have added crossovers to suppress stray microwave chip modes by tying the ground planes together with low impedance connections. Otherwise, the many control wires in our chip could lead to segmentation of the ground plane, and the appearance of parasitic slotline modes [2]. The device is made in a five-step deposition process, (I) Al deposition and control wiring etch, (II) crossover dielectric deposition, (III) crossover Al deposition, (IV) Qubit capacitor and resonator etch, (V) Josephson junction deposition. The qubit capacitor, ground plane, readout resonators, and control wiring are made using molecular beam epitaxy (MBE)-grown Al on sapphire [3]. The control wiring is patterned using lithography and etching with a $BCl_3/Cl_2$ reactive ion etch. A 200 nm thick layer of $SiO_2$ for the crossover dielectric is deposited in an e-beam evaporator, followed by lift-off. We fabricate crossovers on all the control wiring, using a $SiO_2$ dielectric that has a non-negligible loss tangent. An in-situ Ar ion mill is used to remove the native $AlO_x$ insulator, after which a 200 nm Al layer for the crossover wiring is deposited in an e-beam evaporator, followed by lift-off. We used 0.9 $\mu$m i-line photoresist, lift-off is done in N-methyl-2-pyrrolidone at 80°C. A second $BCl_3/Cl_2$ etch is used to define the qubit capacitor and resonators; this step is separate from the wiring etch to prevent the sensitive capacitor from seeing extra processing. Lastly, we use electron beam lithography, an in-situ Ar ion mill, and double angle shadow evaporation to deposit the Josephson junctions, in a final lift-off process. See Ref. [1] for details.

TABLE S1: Qubit frequencies and nonlinearities ($f_{21} - f_{10}$) at the zero flux bias (degeneracy point) and coupling strengths in MHz. The coupling strength is measured at frequencies between 4.2 and 4.7 GHz. We find a typical next-nearest neighbour coupling of $g/2\pi = 1.3$ MHz, consistent with microwave circuit simulations.

| qubits | $Q_0$ | $Q_1$ | $Q_2$ | $Q_3$ | $Q_4$ |
|---|---|---|---|---|---|
| $f_{10}$ | 5805 | 5238 | 5780 | 5060 | 5696 |
| nonlinearity | -217 | -226 | -214 | -212 | -223 |
| $g_{01}/2\pi$ (4.22 GHz) | 27.7 | | | | |
| $g_{12}/2\pi$ (4.70 GHz) | | 30.8 | | | |
| $g_{23}/2\pi$ (4.66 GHz) | | | 30.4 | | |
| $g_{34}/2\pi$ (4.65 GHz) | | | | 30.9 | |

### Coherence Times

Energy relaxation times $T_1$ of all qubits are shown in Fig. S1, measured over a frequency range from 4 to 6 GHz. We find typical $T_1$ values between 20 and 40 $\mu$s. Variations in $T_1$ arise predominantly from the qubit interacting incoherently with weakly coupled two-level defects, as discussed in Ref. [1]. In this previous work we found that larger area (with longer and wider legs) Xmon qubits showed higher $T_1$ values as well as large, frequency-specific suppressions in the energy coherence: for certain frequencies the $T_1$ would decrease to values below 10 $\mu$s. We attribute these large suppressions to chip modes, arising from imbalances in the microwave control lines, to which the larger Xmon geometries can couple more strongly. The data in Fig. S1 exhibit fewer of such suppressions; we believe that this improvement is due to the addition of crossovers.

We have investigated the Ramsey dephasing times versus frequency for qubit $Q_1$. The Ramsey decay envelope is measured by phase tomography (see Ref. [1]) and fitted to the function $\exp[-t/T_{\phi,1} - (t/T_{\phi,2})^2]$. Slow, Gaussian dephasing is captured in $T_{\phi,1}$, and fast dephasing, from white noise as well as energy relaxation, is captured in $T_{\phi,2}$. Typical dephasing times are plotted in Fig. S2. We find a fast dephasing time on the order of 10 $\mu$s; this value is below the energy coherence time, and may be due to white noise from the room temperature control electronics. The slow, Gaussian dephasing times are consistent with a $1/f$-spectrum with a spectral density of $S_\Phi(1\ \text{Hz}) = 1.1\ \mu\Phi_0/\sqrt{\text{Hz}}$.

### Qubit Frequencies and Coupling

Qubit frequencies and nearest neighbour coupling strengths are listed in Table S1.

### Z Crosstalk

We measure a crosstalk between the frequency Z control lines and qubits that is small, approximately $1-2\%$. After adding compensation pulses to orthonormalise the control, we find a remnant crosstalk of below $10^{-4}$. The crosstalk matrix $M_\Phi$ is shown below, defined as: $\Phi_{\text{actual}} = M_\Phi \Phi_{\text{ideal}}$, with $\Phi$ the flux threaded through each qubit's superconducting quantum interference device (SQUID) loop.

$$M_\Phi = \begin{pmatrix} 1.000 & -0.023 & -0.014 & -0.009 & -0.006 \\ 0.019 & 1.000 & -0.022 & -0.011 & -0.007 \\ 0.017 & 0.000 & 1.000 & -0.016 & -0.009 \\ 0.016 & 0.008 & -0.015 & 1.000 & -0.014 \\ 0.013 & 0.014 & -0.016 & -0.010 & 1.000 \end{pmatrix}$$

## EXPERIMENTAL SETUP

The wiring diagram and circuit components are shown in Fig. S3.

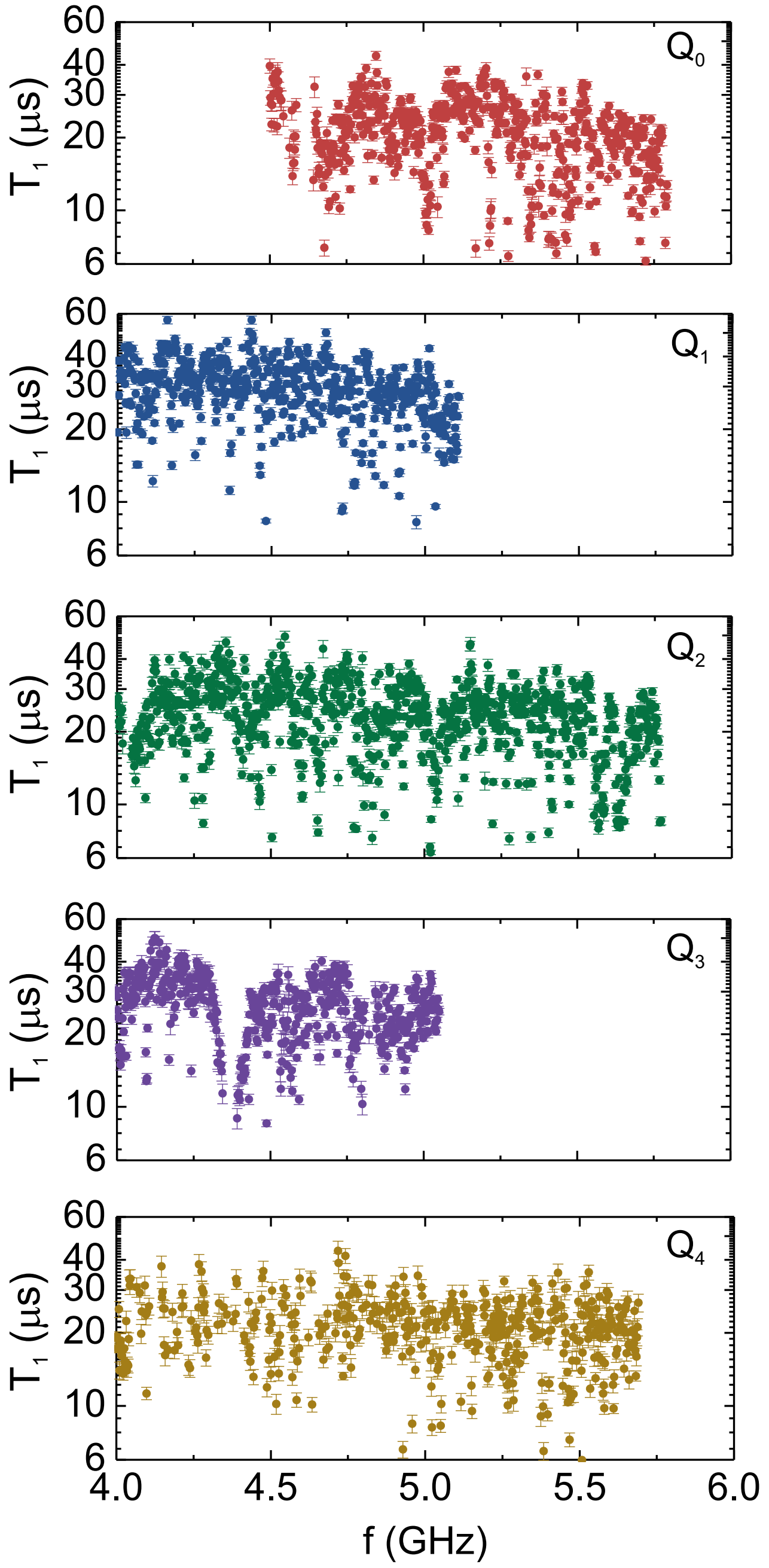

FIG. S1: **Energy Relaxation for Xmon Qubits.** Frequency dependence of $T_1$ for all qubits. The frequency step size is 2 MHz. The values for $T_1$ are generally in the 20-40$\mu$s range. The depression at 4.36 GHz in qubit $Q_3$ is due to a coherently coupled junction defect.

## FLATTENING THE Z RESPONSE

Imperfections in the frequency control wiring can cause ripples after a pulse. Left unchecked, these can affect gate fidelity significantly, appearing as single qubit phase errors, see Fig. S4. We employ a two-step procedure to correct for these non-idealities. We first calibrate the room temperature electronics by measuring the unit step (Heaviside step) response at the output of the Z control board.

With the board response corrected by deconvolution, we measure the qubit phase as a function of time $\Delta\tau$ after the end of a unit step. This probes the transfer function of the fridge wiring, contact pads and on-chip control lines. When no unit step is applied, the X/2 pulse rotates the qubit state onto the Y axis. When applying a unit step, deviations in frequency will cause the Bloch sphere vector to deviate from the Y axis. A subsequent Y/2 pulse will make this apparent in the measured excited state probability. We note that this measurement is first order sensitive to small deviations – the difference in probability denotes the phase deviation ($\Delta\phi \approx \Delta P_{|1\rangle}$) – whereas Ramsey and quantum state tomography are second order sensitive: The $\pi/2$ pulses used in tomography project the state onto the Z axis, thus the reconstruction of the phase or state is done from probabilities ($P \approx 1 - \Delta\phi^2/2$) which are second order sensitive to $\phi$.

We find that the transfer function can be described by an exponential response with two timescales. Typical values are 100 ns and 5 ns. The longer timescale is consistent with the $L/R$ time arising from the bias tee, with $L \approx 6$ $\mu$H and $R = 50$ $\Omega$. We believe that the short timescale arises from reflections. The impulse response of an imperfect wire with reflection $r$, placed time $T$ away from the wire's end is $H(\omega) = 1 - r + \sum_{k=1}^{\infty} r^k \exp(-2ik\omega T)$; at low frequencies this can be approximated by the impulse response function $h(t) \propto \exp(-t/2aT)u(t)$. Assuming reflection coefficients on the order of -10 dB and round trip times $2T$ between qubit and mixing plate electronics on the order several ns, the effective decay time $2rT$ is on the order of a few ns.

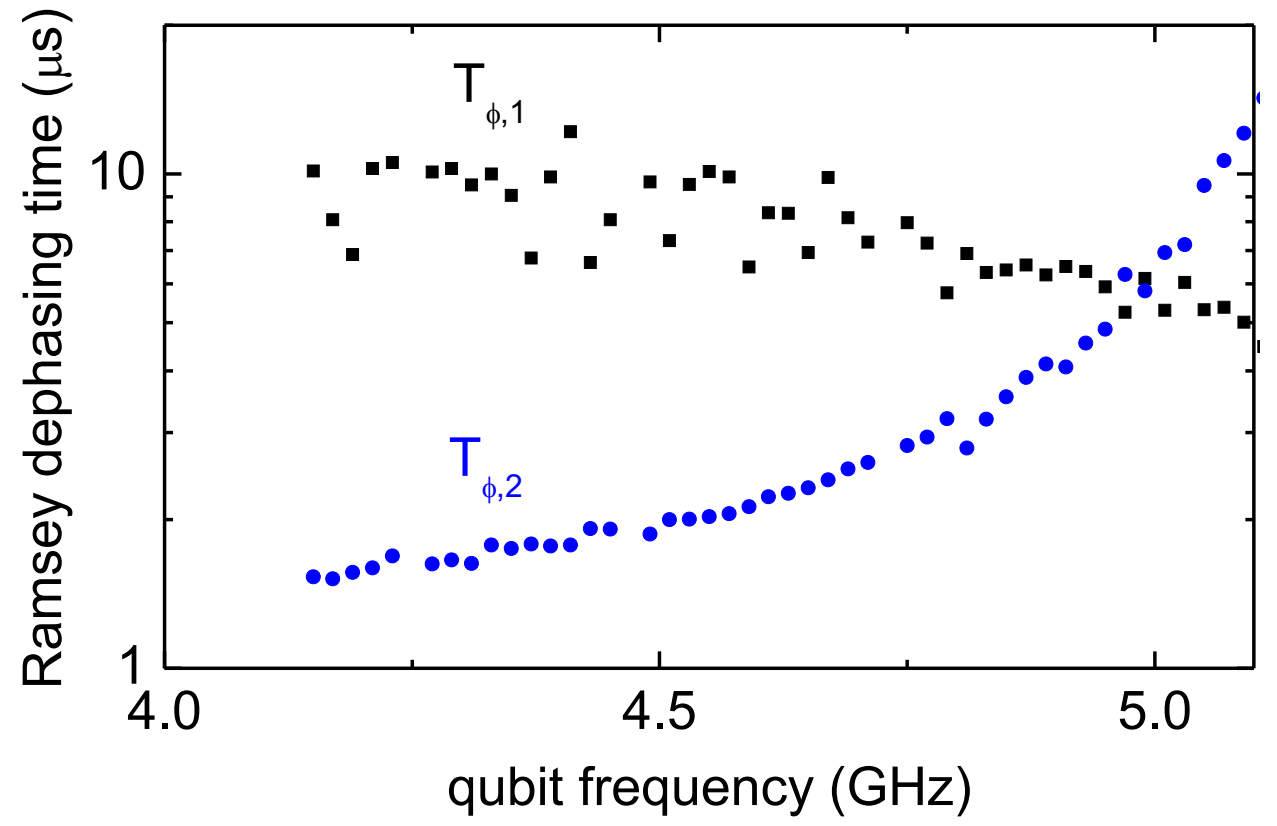

FIG. S2: **Ramsey dephasing.** Frequency dependence of the slow ($T_{\phi,1}$) and fast ($T_{\phi,2}$) Ramsey dephasing times of qubit $Q_1$. Flux bias points $\Phi/\Phi_0$ range from 0.1 to 0.28, and $\delta f_{10}/\delta(\Phi/\Phi_0)$ range from -16 to -50 GHz.

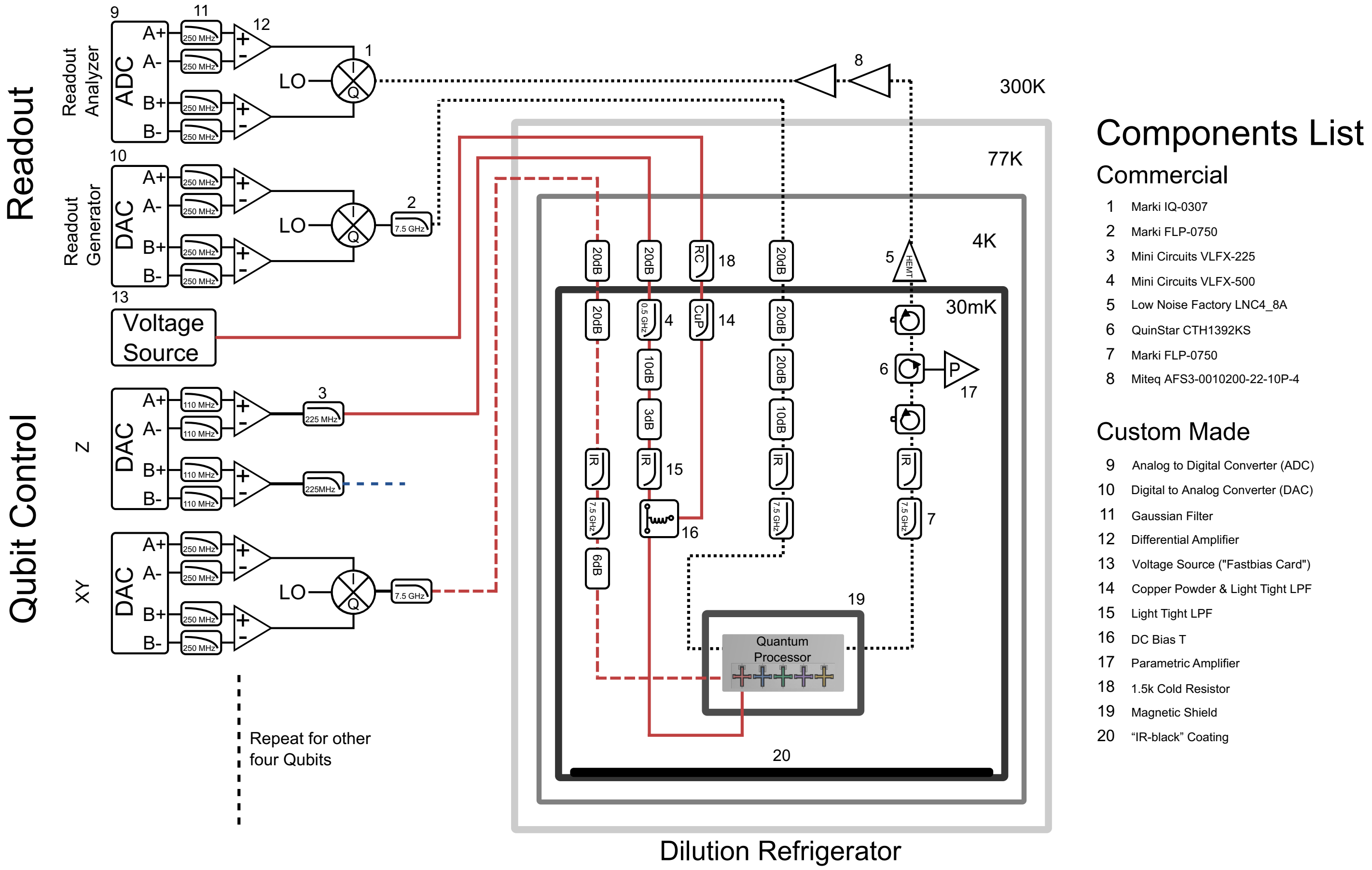

FIG. S3: **Electronics and Control Wiring.** Diagram detailing all of the control electronics, control wiring, and filtering for the experimental setup. Each qubit uses one digital to analog converter (DAC) channel for each of the X, Y, and Z rotations. Additionally, we use a DC bias tee to connect a voltage source to each qubit frequency control line to give a static frequency offset. All five qubits are read out using frequency-domain multiplexing on a single measurement line. The readout DAC generates five measurement tones at the distinct frequencies corresponding to each qubit's readout resonator. The signal is amplified by a wideband parametric amplifier [Mutus, J. *et al.* in preparation], a high electron mobility transistor (HEMT), and room temperature amplifiers before demodulation and state discrimination by the analog to digital converter (ADC). All control wires go through various stages of attenuation and filtering to prevent unwanted signals from disturbing the quantum processor.

With the corrections in place, by deconvolving both the board response and fridge wiring, remnant control pulse ripples are suppressed to below $10^{-4}$: We find qubit phase deviations consistent with a 30 kHz drift after applying a 0.5 GHz detuning step pulse, see Fig. S4. The calibrations discussed above are key for obtaining accurate CZ gates.

## SINGLE QUBIT AND TWO-QUBIT GATE FIDELITIES OF ALL QUBITS

A comprehensive listing of all single qubit gate fidelities of all qubits is shown in Table S2, the gate durations are in Table S3. A listing of all CZ gate fidelities can be found in Table S4.

## VERIFYING EXPERIMENTAL FIDELITIES ARE AT THE SURFACE CODE THRESHOLD

Nominally, the threshold fidelity of the surface code is 0.99 [7], provided one assumes there is no leakage, the two-qubit interaction is the dominant source of error, and gates can be performed perfectly in parallel. The physical device described in this work has complex behavior outside these assumptions, necessitating a device-specific calculation of the surface code threshold fidelity.

When a CZ gate is applied, no qubit neighbouring either of the qubits involved in the CZ can be involved in their own CZ gate. We have devised a 16 step CZ application pattern that accounts for these parallelism constraints and still measures all stabilizers. The longest measured CZ time of 45 ns will be used. Furthermore, the CZ gate, which is always applied between one measurement qubit and one data qubit, has a small amount of leakage

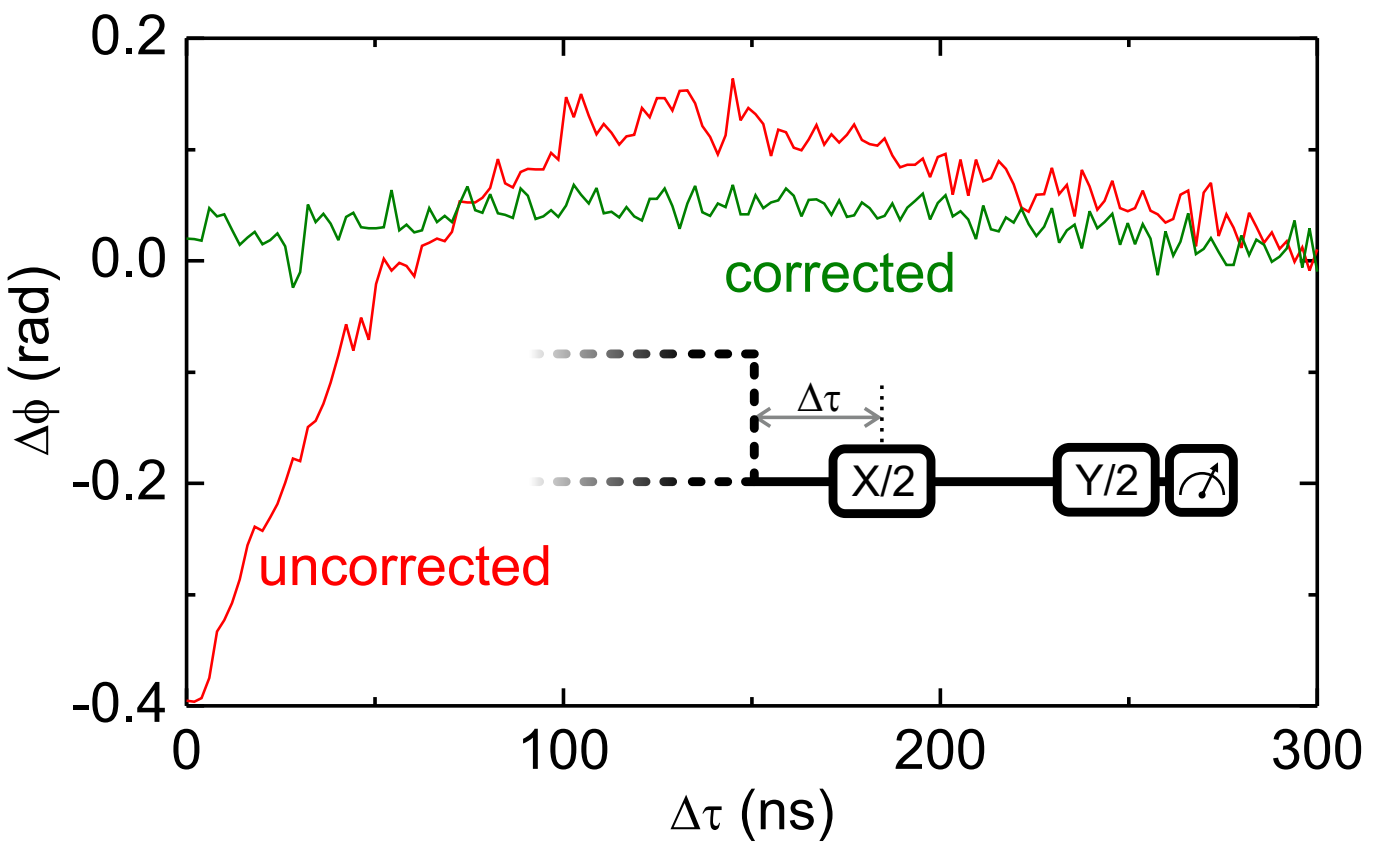

FIG. S4: **Control pulse ripple.** Qubit phase response to a unit step (amplitude: 0.5 GHz) applied to the frequency control line, with and without correction. The pulse sequence is shown in the inset, with the dashed line representing the unit step. With correction, a phase drift of 0.03 rad in 150 ns is observed, consistent with a remnant control pulse ripple of 30 kHz.

($< 0.2\%$) on the measurement qubit, but practically negligible leakage on the data qubit. We shall neglect this small amount of measurement qubit leakage. Methods of coping with leakage in topological codes are known [8].

TABLE S2: Single qubit gate fidelities for all qubits, determined by Clifford-based randomised benchmarking. Averaged over all gates and all qubits we find an average fidelity of 0.9992. The standard deviation is typically $5 \cdot 10^{-5}$. The gate times are between 10 and 20 ns, see Table S3, except for the composite gates H and 2T, which are twice as long. The idle is as long as the shortest microwave gate (12 ns to 20 ns).

| gates | $Q_0$ | $Q_1$ | $Q_2$ | $Q_3$ | $Q_4$ |
|---|---|---|---|---|---|
| I | 0.9990 | 0.9996 | 0.9995 | 0.9994 | 0.9991 |
| X | 0.9992 | 0.9996 | 0.9992 | 0.9991 | 0.9991 |
| Y | 0.9991 | 0.9995 | 0.9993 | 0.9992 | 0.9991 |
| X/2 | 0.9992 | 0.9993 | 0.9993 | 0.9994 | 0.9993 |
| Y/2 | 0.9991 | 0.9993 | 0.9995 | 0.9994 | 0.9994 |
| -X | 0.9991 | 0.9995 | 0.9992 | 0.9989 | 0.9991 |
| -Y | 0.9991 | 0.9995 | 0.9991 | 0.9987 | 0.9991 |
| -X/2 | 0.9991 | 0.9992 | 0.9993 | 0.9990 | 0.9995 |
| -Y/2 | 0.9991 | 0.9992 | 0.9995 | 0.9990 | 0.9994 |
| H | 0.9986 | 0.9986 | 0.9991 | 0.9981 | 0.9988 |
| Z | 0.9995 | 0.9988 | 0.9994 | 0.9991 | 0.9993 |
| Z/2 | 0.9998 | 0.9991 | 0.9998 | 0.9995 | 0.9996 |
| 2T [a] | | 0.9989 | 0.9994 | 0.9989 | 0.9990 |
| average over gates | 0.9992 | 0.9992 | 0.9994 | 0.9991 | 0.9992 |
| average over qubits | | | 0.9992 | | |

[a] As the T gate is not a Clifford generator, a single recovery gate can not be found when interleaving. This precludes Clifford-based randomised benchmarking of the T gate. To quantify this gate to some extent, we have benchmarked 2T gates, physically implemented by applying two T gates in series. If the gate error is predominantly gate-aspecific, the T gate error is half that of the 2T gate, suggesting that the average T gate fidelity is 0.9995

Measurement with fidelity 0.99 in 200 ns and initialization with fidelity 0.99 in 50 ns will be assumed [Jeffrey, E., *et al.*, in preparation]. Y/2 gates will be used instead of Hadamard gates, with the slowest 20 ns time assumed and an average fidelity of 0.9992 (calculated only from the slower Y/2 gates) assumed. An identity error of 0.05% per 10 ns will be assumed, consistent with experimental data.

Detailed simulations of 5x5, 9x9, and 13x13 qubit arrays with the above parameters have been performed making use of the latest correction techniques [9]. The logical error rate was found to be the same in all cases, justifying our claim of a device with parameters at the surface code threshold.

## CZ GATE ERROR BUDGET

We experimentally measure the three predominant error mechanisms of the CZ gate: 2-state leakage, decoherence and control error. 2-state leakage is measured using the same technique as outlined in [4]. The system is initialised in the $|11\rangle$-state followed by two CZ gates. As the time between these two gates is varied, we measure an interference pattern in the probability where the amplitude is proportional to the $|02\rangle$ state leakage. The error is given by $\Delta P/4$ [4]. We also see additional interference patterns that come from imperfect $|11\rangle$ preparation at the beginning of the sequence. We note that leakage occurs predominantly in the qubit which undergoes the frequency trajectory.

TABLE S3: Single qubit gate times in ns.

| gates | $Q_0$ | $Q_1$ | $Q_2$ | $Q_3$ | $Q_4$ |
|---|---|---|---|---|---|
| XY axes $\pi$ rotations | 20 | 20 | 12 | 18 | 12 |
| XY axes $\pi/2$ rotations | 20 | 20 | 12 | 12 | 12 |
| Z axis $\pi$, $\pi/2$, $\pi/4$ rotations | 10 | 10 | 10 | 10 | 10 |
| I | 20 | 20 | 12 | 12 | 12 |
| H | 40 | 40 | 24 | 30 | 24 |
| 2T | 20 | 20 | 20 | 20 | 20 |

TABLE S4: CZ gate fidelities for all qubit pairs, determined by Clifford-based randomised benchmarking. Gate times are between 38 and 45 ns; $Q_0$-$Q_1$: 45 ns, $Q_1$-$Q_2$: 43 ns, $Q_2$-$Q_3$: 43 ns, $Q_3$-$Q_4$: 38 ns.

| qubits | $Q_0$ | $Q_1$ | $Q_2$ | $Q_3$ | $Q_4$ |
|---|---|---|---|---|---|
| $CZ_{Q_0-Q_1}$ | $0.9924 \pm 0.0005$ | | | | |
| $CZ_{Q_1-Q_2}$ | | $0.9936 \pm 0.0004$ | | | |
| $CZ_{Q_2-Q_3}$ | | | $0.9944 \pm 0.0005$ | | |
| $CZ_{Q_3-Q_4}$ | | | | $0.9900 \pm 0.0006$ | |

TABLE S5: CZ gate error budget, including the contribution to the total error in percent.

| | | |
|---|---|---|
| Decoherence (55%) | $Q_2$ | $\geq 0.0017$ (24%) |
| | $Q_3$ | 0.0022 (31%) |
| Control (45%) | single qubit phase error | $\leq 0.0017$ (24%) |
| | state leakage | 0.0015 (21%) |

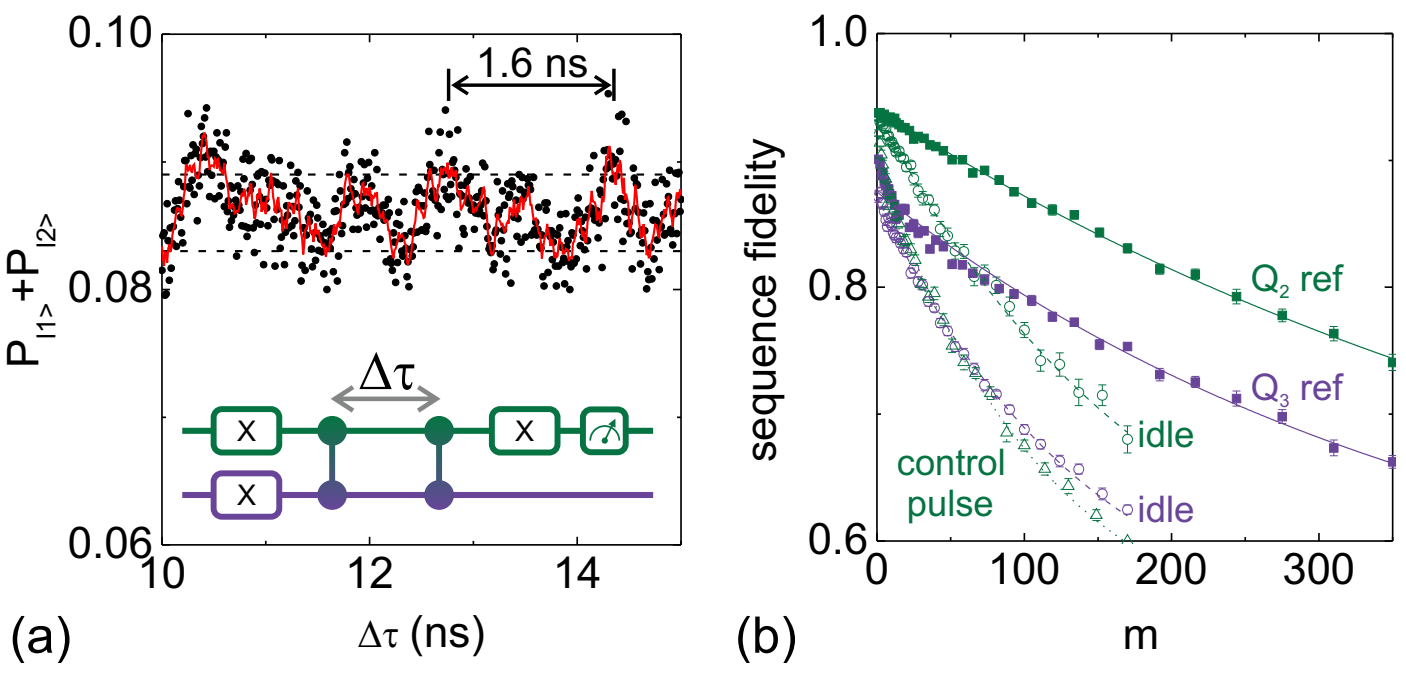

FIG. S5: **CZ error budget.** (a) In the Ramsey error filter technique an interference pattern arises in the measured probability (black dots) whose magnitude is proportional to the $|02\rangle$-state leakage. The data are smoothed (red) for enhanced visibility. The frequency of these oscillations (indicated by the arrow, 1.6 ns) is the idling frequency difference between qubits (800 MHz), minus the nonlinearity (200 MHz) as we are measuring the crossing between $|11\rangle$ and $|02\rangle$. Other frequencies are believed to arise from improper $|11\rangle$ state preparation. (b) Randomised benchmarking sequence fidelity for qubits $Q_2$ and $Q_3$. Decoherence is quantified by idling for the same duration as the CZ gate. Controls error can be identified by applying the control pulse on $Q_2$, without doing a full CZ by detuning $Q_3$. For the randomised benchmarking data of the CZ (not shown): $r_{\rm ref} = 0.0198$ and $r_{\rm CZ} = 0.0269$.

We measure the decoherence contribution from each qubit by performing interleaved randomised benchmarking with an idle of the same duration as the CZ gate. The contribution to error from the waveform is measured by interleaved randomised benchmarking on the waveform for the CZ gate alone, with a slightly lower amplitude to avoid interactions with the other qubit. We treat this as a single-qubit phase gate. With the idle error measured, we can separate out decoherence and single qubit phase error. Because we are detuning the qubit down in frequency to a part of the spectrum where it is more sensitive to flux noise, inducing more dephasing, the single qubit phase error is an upper bound. With these experiments we can construct an error budget for all of the dominant error mechanisms, as seen in Table S5.

## QUANTIFYING XY CONTROL CROSSTALK USING SIMULTANEOUS RANDOMISED BENCHMARKING

Addressability, the ability to individually control a single qubit without affecting neighbouring qubits, is of great importance when building a multi-qubit system. In our five Xmon qubit processor the addressability is mostly compromised in three ways: Z control crosstalk, microwave XY control crosstalk, and off-resonant qubit-qubit coupling. Z crosstalk can be reduced to below the $10^{-4}$ level. Microwave XY crosstalk becomes a problem if a qubit's control pulses perform rotations on a neighbouring qubit. Off-resonant qubit-qubit coupling will very slowly perform a CZ gate between the qubits, potentially causing unwanted phase shifts with rate $\Omega_{ZZ}$,

$$\Omega_{ZZ} = -\frac{2g^2(\eta_1+\eta_2)}{(\Delta-\eta_1)(\Delta+\eta_2)}, \tag{S1}$$

with $\eta_1$ and $\eta_2$ the qubit nonlinearities, and $\Delta$ the difference in qubit frequencies.

We performed crosstalk characterisation on nearest neighbour and next-nearest neighbour qubits. Nearest neighbours are far detuned ($>$ 800 MHz), hence the microwave XY crosstalk is expected to be negligible, but the off-resonant CZ interaction may be non-negligible. Next-nearest neighbors have a much smaller coupling ($g$ =1.3 MHz), but are only detuned by 100-400 MHz; hence both the off-resonant CZ as well as microwave XY crosstalk may be detrimental. We investigate these mechanisms by using the simultaneous randomised benchmarking techniques outlined in [6]. We can single out errors that come from poor addressability by performing randomised benchmarking on each qubit individually, and operating both qubits simultaneously.

The randomised benchmarking data are shown in Fig. S6. We can determine the effect of controlling qubit $Q_3$ on $Q_2$, by first benchmarking qubit $Q_2$ individually ($I \otimes C_1$, green open squares), and benchmarking both qubit $Q_2$ and $Q_3$ simultaneously, and tracing out the contribution of $Q_3$ ($C_1 \otimes C_1$, green full squares). The decay for both traces is virtually indistinguishable, the added error is below $10^{-4}$. Likewise, we find that the effect on $Q_3$ of controlling $Q_2$ simultaneously leads to an added error per Clifford of $2 \cdot 10^{-4}$. For next nearest neighbours, we find added errors per Clifford of $1 \cdot 10^{-4}$ and $2 \cdot 10^{-4}$. For both the nearest neighbour and next-nearest neighbours the added error of operating them simultaneously is $< 2 \cdot 10^{-4}$. We conclude that XY control crosstalk is a minor error mechanism, enabling a high degree of addressability in this architecture.

## GENERATION OF THE CLIFFORD GROUPS

### Single qubit Clifford group $C_1$

The single qubit Clifford group $C_1$ is the group of 24 rotations which preserve the octahedron in the Bloch sphere. We implement the group using microwave pulses only, decomposed into rotations around the X and Y axes using the generators: {I, ±X/2, ±Y/2, ±X, ±Y }, as summarised in Table S6. The average number of single qubit gates per single qubit Clifford is 1.875.

### Two qubit Clifford group $C_2$

Using the single qubit Cliffords, we can construct the two qubit Clifford group $C_2$ following Ref. [5]. This group has four classes: the single qubit class, the CNOT-like class, the iSWAP-like class, and the SWAP-like class. The CNOT and SWAP-like

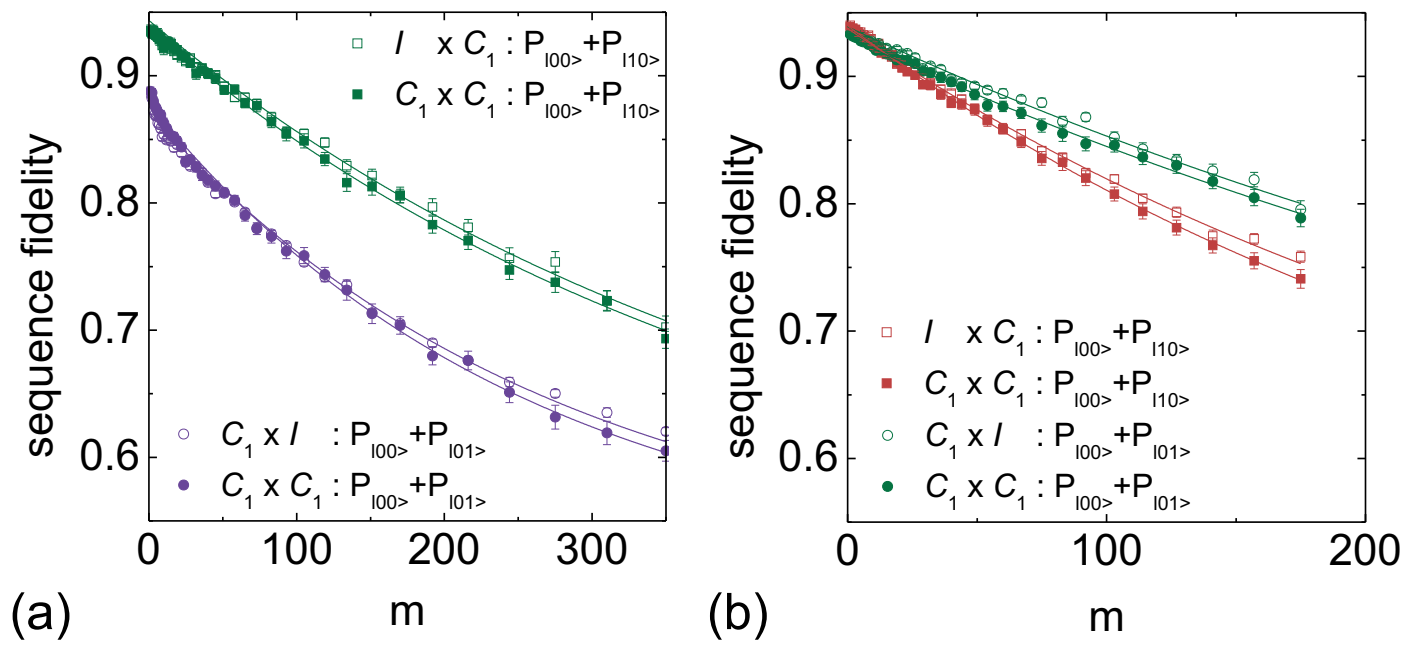

FIG. S6: **Simultaneous randomised benchmarking of nearest and next nearest neighbours.** (a) Benchmarking the effect of $Q_2$ on $Q_3$ and vice-versa ($f_{Q_2} = 5.72$ GHz, $f_{Q_3} = 4.67$ GHz). The sequence fidelities are shown for operating $Q_2$ individually ($I \otimes C_1$, green open squares), $Q_3$ individually ($C_1 \otimes I$, purple open circles), and $Q_2$ and $Q_3$ simultaneously ($C_1 \otimes C_1$, full symbols). By tracing out one qubit, its effect on the other qubit becomes apparent: the errors per Clifford are: $r_{Q2}$=0.0011, $r_{Q2|Q3}$=0.0012, $r_{Q3}$=0.0018, $r_{Q3|Q2}$=0.0020. (b) Benchmarking of $Q_0$ and $Q_2$ ($f_{Q_0} = 5.30$ GHz, $f_{Q_2} = 5.72$ GHz). The errors per Clifford are: $r_{Q0}$=0.0016, $r_{Q0|Q2}$=0.0018, $r_{Q2}$=0.0011, $r_{Q2|Q0}$=0.0011. Note that the errors per Clifford are consistent with the average gate fidelities in Table S2: for $Q_2$, the average gate fidelity is $1 - r_{Q2}/1.875$=0.9994. Coupling strengths can be found in Table S1.

class are terminated with a gate from the 3-element group $S_1$, as described in Table S7. The single qubit class has $24^2 = 576$

TABLE S6: The 24 single qubit Cliffords written in terms of the physical microwave gates applied in time. The Paulis and $2\pi/3$ rotations form the tetrahedron symmetry group.

| Single qubit Cliffords | |
|---|---|
| Paulis | I |
| | X |
| | Y |
| | Y, X |
| $2\pi/3$ rotations | X/2, Y/2 |
| | X/2, -Y/2 |
| | -X/2, Y/2 |
| | -X/2, -Y/2 |
| | Y/2, X/2 |
| | Y/2, -X/2 |
| | -Y/2, X/2 |
| | -Y/2, -X/2 |
| $\pi/2$ rotations | X/2 |
| | -X/2 |
| | Y/2 |
| | -Y/2 |
| | -X/2, Y/2, X/2 |
| | -X/2, -Y/2, X/2 |
| Hadamard-like | X, Y/2 |
| | X, -Y/2 |
| | Y, X/2 |
| | Y, -X/2 |
| | X/2, Y/2, X/2 |
| | -X/2, Y/2, -X/2 |

elements:

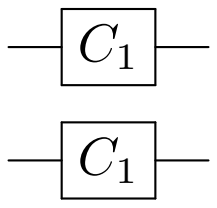

The CNOT-like class has $24^2 \times 3^2 = 5184$ elements.

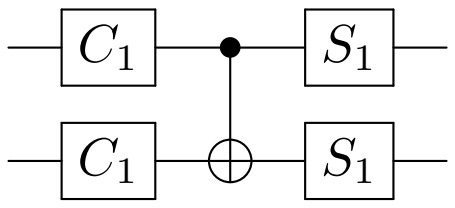

The iSWAP-like class also has 5184 elements,

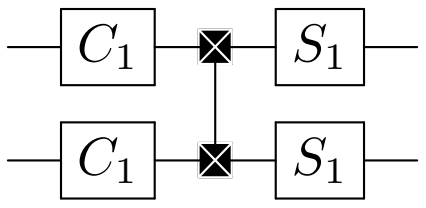

Finally the SWAP-like class, with 576 elements, is given by bringing the full size of the two-qubit Clifford group to 11520.

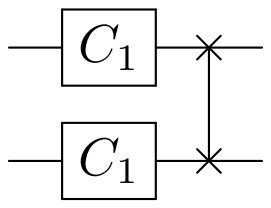

Here, we rewrite the two-qubit Cliffords in terms of the CZ entangling gate. We rewrite the CNOT, iSWAP and SWAP in terms of the CZ:

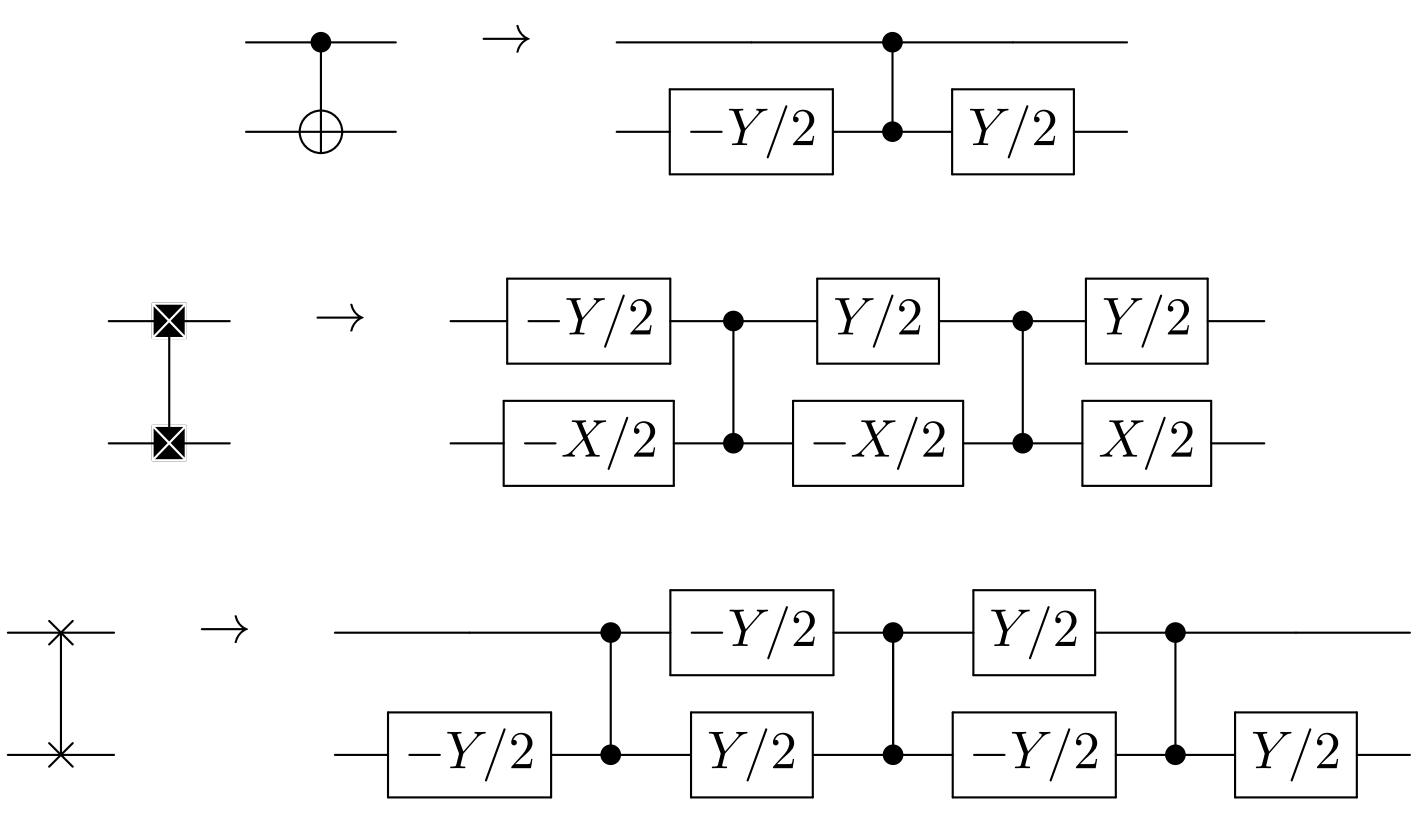

As the single qubit gates preceeding the entangling operation (CZ gate) can be absorbed into $C_1$, and the final single qubit

TABLE S7: The $S_1$ sets written in terms of physical gates in time; these are elements of the single qubit Clifford group, and therefore physically implemented in the same way.

| | |
|---|---|
| $S_1$ | I |
| | Y/2, X/2 |
| | -X/2, -Y/2 |
| $S_1^{X/2}$ | X/2 |
| | X/2, Y/2, X/2 |
| | -Y/2 |
| $S_1^{Y/2}$ | Y/2 |
| | Y, X/2 |
| | -X/2, -Y/2, X/2 |

gates can be absorbed into $S_1$ (see Table S7), we have for the CNOT-like class,

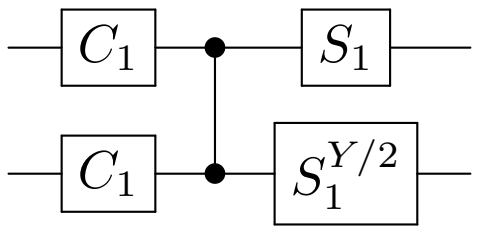

the iSWAP-like class,

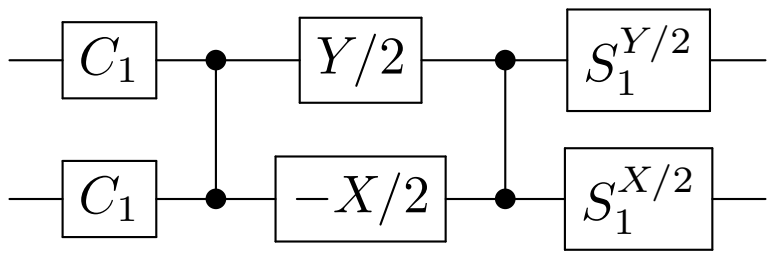

and the SWAP-like class,

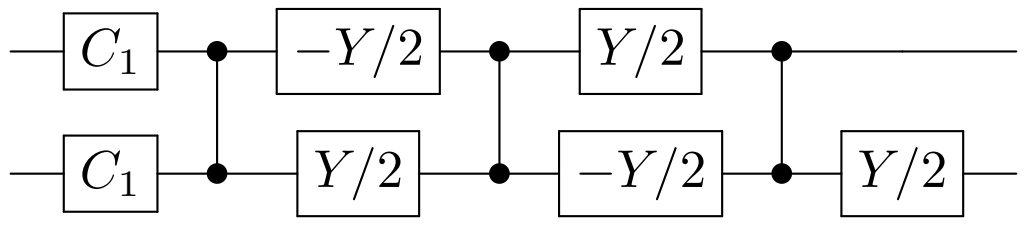

The average number of gates for $C_2$ is 1.5 CZ gates and 8.25 single qubit gates. For the idle, we wait as long as the shortest single qubit gate. For a single qubit gate time of 20 ns and a CZ gate time of 40 ns, the average duration of $C_1$ is 37.5 ns, and $C_2$ is 160 ns.

The Clifford group is a 2-design. A set of unitaries $\{U_k\}_{k=1}^{K}$ is a 2-design if and only if [12]

$$\sum_{k,k'=1}^{K} |\mathrm{Tr}\left(U_{k'}^{\dagger} U_k\right)|^4 / K^2 = 2. \tag{S2}$$

As a consistency check, we have verified that the single and two-qubit Cliffords we generate are indeed a 2-design with the above equation.

## ESTIMATING THE ERROR PER CLIFFORD

Here, we connect the error per Clifford $r$ to the errors of the single and two-qubit gates, measured when performing randomised benchmarking. This shows the physical significance of the error per Clifford, and is an important self-consistency check. We can give an estimate for the error per Clifford by determining the average number of single and two-qubit gates that go into a Clifford, combined with the single and two-qubit gate fidelities that we measure using interleaved randomised benchmarking. We assume that all the gates have low enough error such that adding error when composing gates is a good approximation.

### Single qubit Clifford group $C_1$

There are 45 single qubit gates used across 24 Cliffords. With the assumption that all single gates have the same error, the average error per Clifford is

$$r_{C_1} = 1.875 r_{\mathrm{SQ}}, \tag{S3}$$

with $r_{\mathrm{SQ}}$ the average single-qubit gate error.

### Two qubit Clifford group $C_2$

The four classes of two-qubit Cliffords are composed from the two-qubit CZ gate, and the single-qubit gate sets $C_1$, $S_1$, $S_1^{Y/2}$, and $S_1^{X/2}$. The respective errors are given by: $r_{S_1} = 5r_{\mathrm{SQ}}/3$, $r_{S_1^{X/2}} = 5r_{\mathrm{SQ}}/3$, $r_{S_1^{Y/2}} = 2r_{\mathrm{SQ}}$.

We now derive the average gate composition for the two-qubit Cliffords. For the single-qubit class:

$$r_{C_1 \otimes C_1} = \frac{90}{24} r_{\mathrm{SQ}}. \tag{S4}$$

CNOT-like class:

$$r_{\mathrm{CNOT}} = r_{\mathrm{CZ}} + \frac{89}{12} r_{\mathrm{SQ}}. \tag{S5}$$

iSWAP-like class:

$$r_{\mathrm{iSWAP}} = 2r_{\mathrm{CZ}} + \frac{113}{12} r_{\mathrm{SQ}}. \tag{S6}$$

SWAP-like class:

$$r_{\mathrm{SWAP}} = 3r_{\mathrm{CZ}} + \frac{35}{4} r_{\mathrm{SQ}}. \tag{S7}$$

The error per Clifford for $C_2$ is then given by

$$r_{C_2} = \frac{576}{11520} r_{C_1 \otimes C_1} + \frac{5184}{11520} r_{\mathrm{CNOT}} + \tag{S8}$$

$$\frac{5184}{11520} r_{\mathrm{iSWAP}} + \frac{576}{11520} r_{\mathrm{SWAP}} \tag{S9}$$

$$= \frac{3}{2} r_{\mathrm{CZ}} + \frac{33}{4} r_{\mathrm{SQ}}. \tag{S10}$$

And the error per two-qubit Cliffords interleaved with a CZ is

$$r_{C_2 + \mathrm{CZ}} = \frac{5}{2} r_{\mathrm{CZ}} + \frac{33}{4} r_{\mathrm{SQ}}. \tag{S11}$$

### Comparison to Experiment

Using these simple formulas, we find that our randomised benchmarking data are self-consistent. Using reasonable values of 0.001 and 0.006 for the single and two-qubit gates respectively, we calculate $r_{C_2} = 0.0173$, which is close to the experimental value of $r_{\mathrm{ref}} = 0.0189$ in Fig. 3 in the main Letter; for the interleaved case the calculated value of $r_{C_2 + \mathrm{CZ}} = 0.0233$ is close to the experimental value of 0.0244 as well.

## $N = 5$ GHZ STATE PULSE SEQUENCE

The pulse sequence for the algorithm to construct the five qubit GHZ state is shown in Fig. S7a. We use Hahn spin echoes on idling elements to suppress slow dephasing ($T_{\phi,2}$). The frequency diagram for the qubits is shown in Fig. S7b. Nearest neighbour qubits are detuned by 0.7 to 1.5 GHz, next-nearest neighbours are detuned by 0.4 to 0.5 GHz.

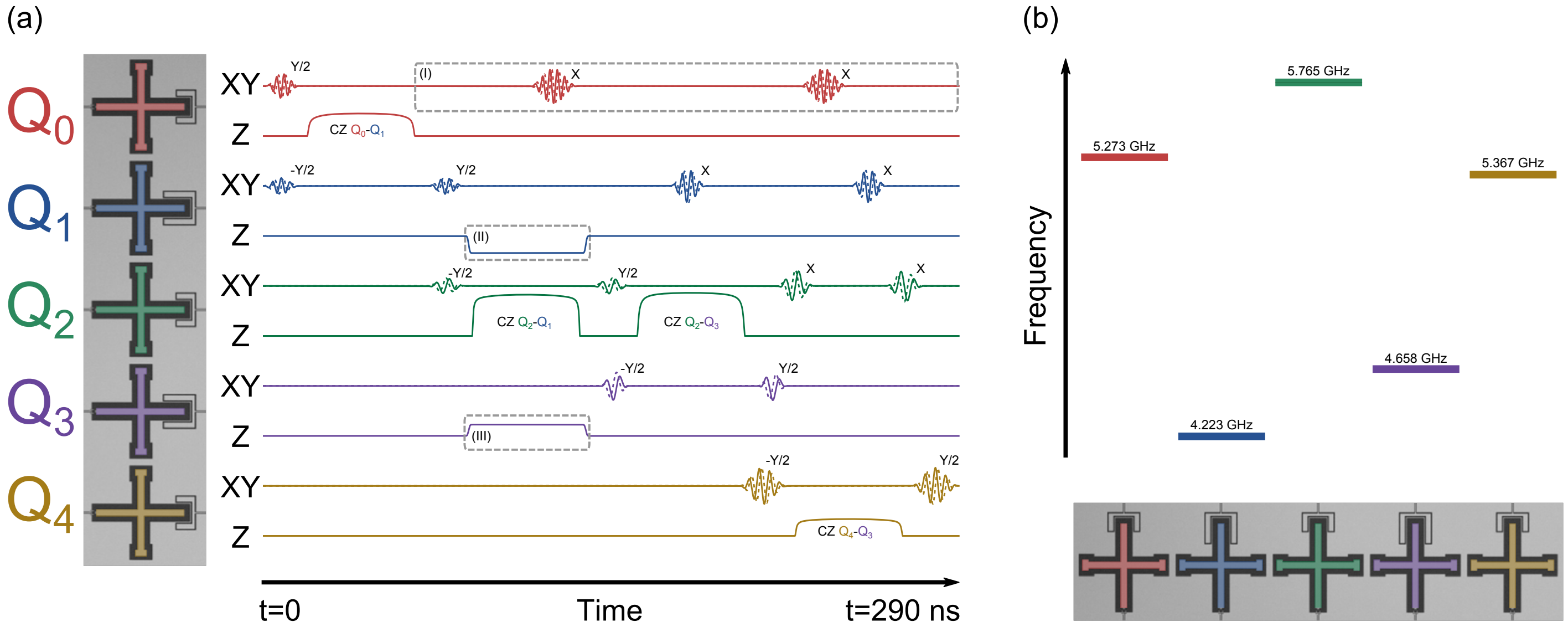

FIG. S7: **Pulse sequence for generating the $N = 5$ GHZ state and frequency diagram.** (a) The control signals for all five qubits. The algorithm consists of a Y/2 pulse on $Q_0$ followed by successive CNOT gates (implemented here with a CZ gate and -Y/2, Y/2 gates applied to the target) on each progressive pair of qubits in the array. The highlighted region (I) shows Hahn spin echo pulses X applied to $Q_0$ to suppress dephasing while idling. Spin echo pulses are also applied to $Q_1$ and $Q_2$. (II) We detune $Q_1$ to bring it closer in frequency to $Q_2$ for the CZ gate. (III) Simultaneously with the $Q_1$-$Q_2$ entangling operation, we detune $Q_3$ away to allow for selective entanglement. (b) The frequency diagram shows the idling frequencies for all qubits, and is one of the operating modes of the quantum processor.

## QUANTUM STATE TOMOGRAPHY

The density matrices of the $N = 2$ Bell and $N = 3, 4, 5$ GHZ states are characterised using quantum state tomography. After state preparation, gates from $\{$ I, X/2, Y/2, X $\}^{\otimes N}$ are applied; with the measured probabilities the state can then be reconstructed. We use quadratic maximum likelihood estimation, using the MATLAB packages SeDuMi and YALMIP, to extract the density matrix while constraining it to be Hermitian, unit trace, and positive semidefinite; the estimation is overconstrained. Non-idealities in measurement and state preparation are suppressed by performing tomography on a zero-time idle [10, 11]. We note that tomography is only "as good as" the tomography pulses, which have an average fidelity above 0.999. Fidelities and uncertainties correspond to the mean and standard deviation of 10 measurements, consisting of $10^4$ ($N = 2, 3$) or $6 \cdot 10^3$ ($N = 4, 5$) repetitions each. The density matrices plotted in the main article are constructed by averaging all measured probabilites, effectively using $10^5$ ($N = 2, 3$) or $6 \cdot 10^4$ ($N = 4, 5$) repetitions.

The imaginary part of the density matrices ($\rho$) is plotted in Fig. S8. The Pauli operator representation is shown in Fig. S9.

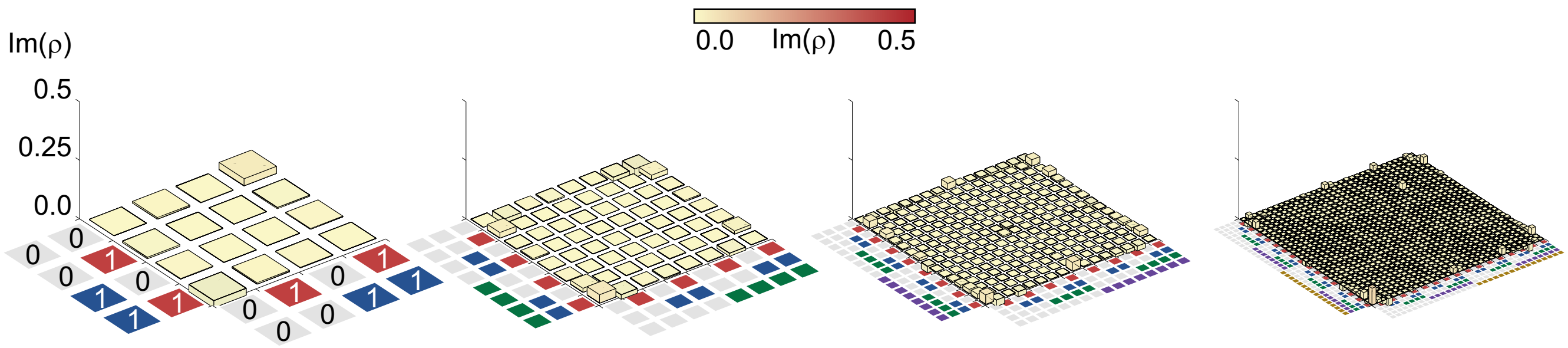

FIG. S8: **Quantum state tomography of the GHZ state: imaginary parts.** Imaginary parts of the density matrices $\rho$ for the $N = 2$ Bell state and the $N = 3, 4$ and $5$ GHZ states. For clarity, the same scale as for the main Letter is used. $|\mathrm{Im}\rho|$ is below 0.03, 0.04, 0.04 and 0.07 for $N = 2$ to 5, respectively.

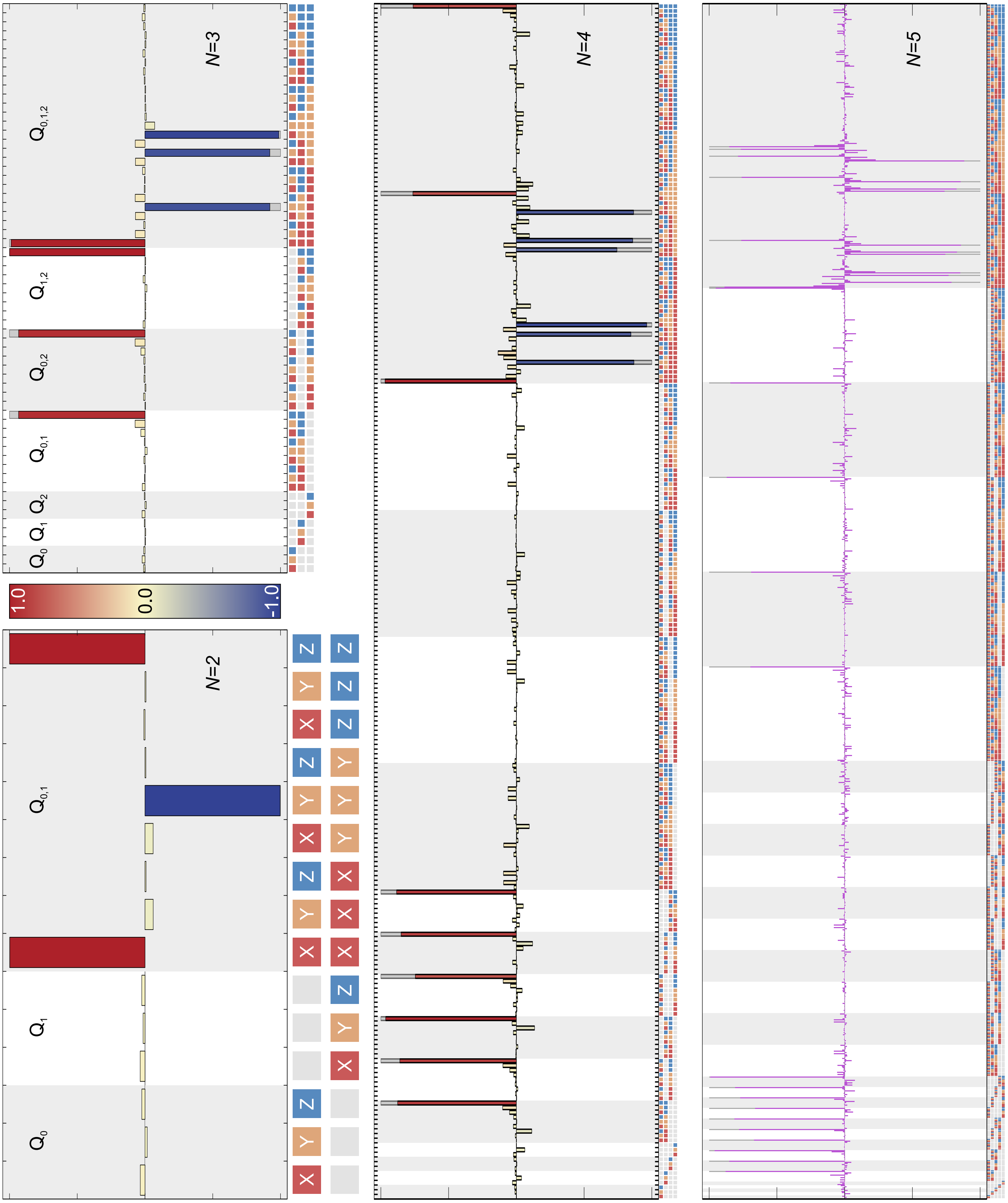

FIG. S9: **Pauli operator representation for the $N = 2$ Bell state and the $N = 3, 4$ and $5$ GHZ states.** The bars show expectation values of combinations of Pauli operators, ideal in grey, experimental values in colour.